# Compliant Self Service Access to Secondary Use Clinical Data at Stanford Medicine

SC Weber, J Pallas, G Olson, D Love, S Malunjkar, S Boosi, E Loh, S Datta, TA Ferris
Research Technology, Technology and Digital Solutions, Stanford Health Care and Stanford School of Medicine, Stanford University

## Abstract

STARR (STAnford Research Repository) is a clinical research support ecosystem that supports basic science research, population health research and translational research at Stanford University. STARR consists of raw and analysis ready multi-modal data, and tools for cohort analysis and self service data access.  STARR data is accessible on secure shared computing systems for ad hoc analysis. Also present is a suite of services on top of STARR, that allow researchers access to complex purpose built data cuts, common data models and software solutions.  **This manuscript is a research resource description** and describes the evolution of "STARR Tools" that are used to offer self-service access to detailed clinical data for research purposes to researchers at Stanford Medicine, along with a framework used to ensure that data acquired via the self-service tools is handled in compliance with all applicable regulations and rules.

## Introduction

In 2003, Stanford Medicine launched a multiyear research and development project to create an integrated, standards-based informatics platform addressing three major barriers to effective CTR: (1) efficient access to clinical data for research purposes; (2) delivery of robust research data management solutions and (3) availability of an enterprise-level system for managing and discovering biospecimen data. The resulting system, launched in 2008, was called STRIDE (Stanford Translational Research Integrated Database Environment) [Lowe2009, Lowe2010], with the stated aim of providing a way at Stanford Medicine to obtain clinical data for secondary use [Xiang2021],  a Health Insurance Portability and Accountability (HIPAA) term that encompasses all non-clinical uses of Protected Health Information (PHI).  Sixteen years later the STARR ecosystem [Callahan 2023, Datta2020] continues to address all three of these barriers, but mostly in ways not originally envisioned. This paper describes the transformative journey that led us to our current suite of support tools, STARR Tools, for clinical research and other secondary uses of hospital patient data at Stanford Medicine.

At the heart of STRIDE was the in-house "Square Table" data model (Figure 1). The cohort and chart review tools interrogate the Square Table data model. While the two hospitals have two different Epic instances that leverage Epic built-in capabilities for seamless patient care, the underlying Epic Clarity data sources needs to be unified in STARR using the master patient index (MPI). The adult hospital provides extensive suite of services to the pediatric hospital, services such as emergency department visits, a set of laboratory services, and radiology imaging data hosting - the patient timeline is unified in STARR using the MPI.

Evolution of STARR since 2017 led to development of other common data models for collaborative research, such as OHDSI OMOP Common Data Model (for both adult and pediatric hospitals), PEDSNet (pediatric hospital only) and PCORNet (adult hospital only). It also led to integration of non-Clarity hospital databases e.g. Philips XPer cardiology data. STARR has non-hospital data sources e.g., Social Security Administration (SSA) Limited Access Death Master File (LADMF). And finally, STARR has integrated diverse data modalities in time such as radiology and opthalmology imaging, cardiology imaging, bedside monitoring and more.

The vision for STARR "self-service" Tools today is to provide self-service access to underlying linked STARR data and modalities. It is designed to offer an intuitive "single pane of glass" view to the complex world of disparate clinical applications at the two hospitals. We envision a future, where the researcher will be able to access the clinical data in a data model of their choice e.g., Square Tables or OMOP, linked to data modalities they need e.g., radiology imaging. Development of specific user facing features in STARR Tools have been research demand driven modulated by availability of resources. Where self-service feature is not available, data access is offered by a team of data service providers. STARR Tools preserves the intended functionality of the cohort and chart review tools that existed in STRIDE while evolving them to meet changing landscape of technology, research and regulatory space. For instance, STARR Tools has recently added capability to allow data download in OMOP data model.

# Methods

## Divergence from the original STRIDE vision

STRIDE was originally built on the Oracle 11g relational database platform using an Entity-Attribute-Value (EAV) model [Brandt2002] to represent object-oriented data structures (entities, roles and acts) derived from the Health Level 7 (HL7) Reference Information Model (RIM) [Lyman2003]. The system included a Master Person Index (MPI) dynamically populated from clinical, research and biospecimen data. HL7 is a healthcare file exchange standard that defines a file format that is widely used for EMR data export.

One of the first departures from our original roadmap was the scrapping of reliance on the RIM. Within 5 years of initial system launch we started the process of transforming our data model to use more conventional database tables, which involved pivoting the attributes into columns and the values into rows. For example, a drug administration record that would have consisted of 10 rows in the EAV model became a single row in a table with 10 columns in our new "square table" data model depicted in Figure 1.

We undertook this transformation primarily for ease of maintenance, as the SQL queries required to manipulate data represented in EAV tables are so complex it was becoming increasingly difficult to keep the system free of bugs.

The next departure from the original roadmap was the adoption of REDCap [Harris2000, Harris2018] as our research data management platform. Stanford's instance of REDCap supports thousands of researchers today and there are over 5000 active data management projects. The use of Stanford REDCap rapidly outstripped the use of STRIDE's home-grown research data management application. In REDCap, researchers build custom research data capture projects on their own, which was both faster and more cost effective for everyone.

We also discovered over time that the needs of biospecimen data management for research have very little overlap with clinical data, and that an entire industry was emerging around biospecimen management tools. With data management now in the purview of the biobanks, we set up periodic data imports of biospecimen availability into STARR, in order to support searching a group of patients (aka a cohort) meeting the enrollment criteria for any given study by their biospecimen availability status. We also created an Application Programming Interface (API) for patient identifier retrieval that is in active use by the current vendor implemented at the Stanford Biobank as described in the API section.

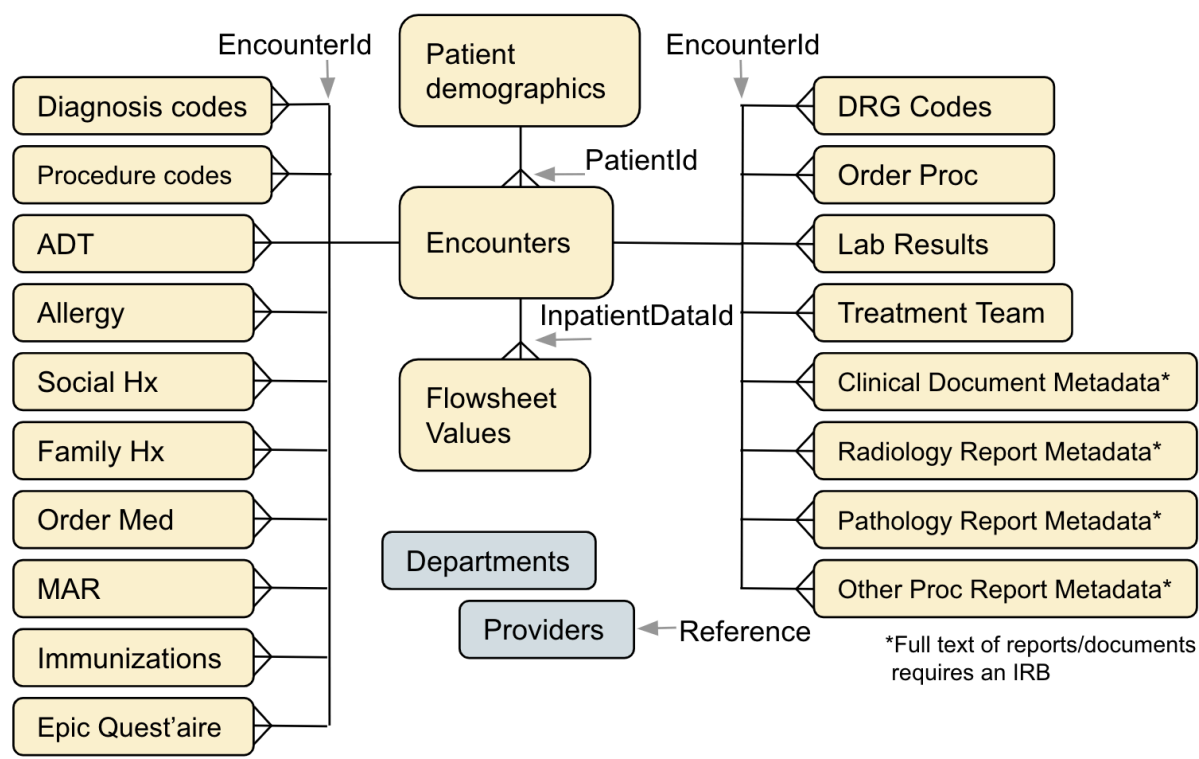

Figure 1: The in-house "Square Table" data model

## Evolution of Cohort and Chart Review Tools

Today, the "Square Table" is the only remaining database from the original STRIDE system (see Figure 2). The first transformative step was moving from relying on HL7 message data to utilizing a single vendor's clinical information system reporting database. Until 2015 Stanford's two hospitals were using different Electronic Medical Record systems (EMRs), with Stanford Health Care using Epic and Stanford Medicine Children's Health using Cerner. In 2015, when Stanford Children's switched to Epic, we were able to implement a nightly Extract Transform and Load (ETL) of Epic Clarity data for both hospitals. HL7 messages are still used, but only for real-time research alerting [Weber2010]. One hospital installed Epic on-premise and the other opted to have Epic host their data, so two distinct ETL processes are used in the data extraction with similar but slightly specialized transformations, to produce a single coherent dataset.

Square Table data model uses multiple terminologies, as all of ICD-9-CM, ICD-10-CM, CPT, and RxNorm [Hernandez2009] are utilized in Epic. Our early focus on mapping idiosyncratic EMR data to these standard terminologies slowly fell into disuse some years after the switch to Epic Clarity. We were only able to phase it out entirely fairly recently, after Epic introduced RxNorm into Clarity and made the mapping reliable.

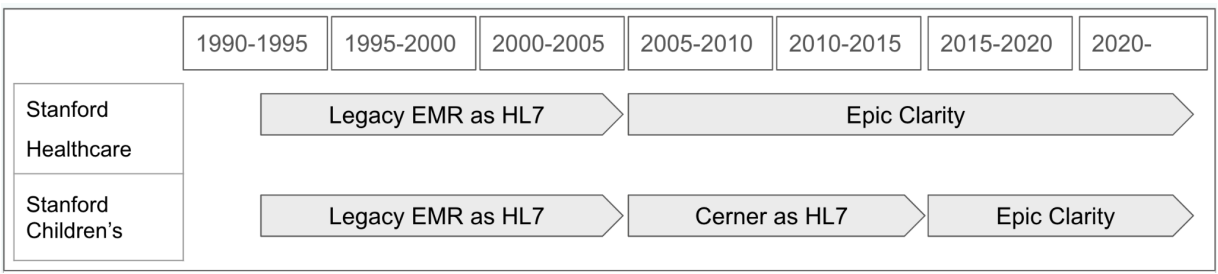

Figure 2: Change in EMR data sources over time at Stanford Medicine

Another transformation we undertook was to switch from a Java Swing "thick client" to a modern web application running entirely in a browser. Our users were expressing frustration at having to authenticate twice, once in the web browser to launch the application, and a second time in order to log into the application. At the same time Java Swing was waning rapidly in popularity, as browsers were being released with built-in capabilities to act as a platform for running web applications. These new powerful web browsers spawned an explosion of competing web application development frameworks. After evaluating multiple options we chose Google Web Toolkit (GWT) as the new framework, a decision which has proven surprisingly resilient. GWT, while no longer a popular choice for new web application development, continues to be sufficiently supported by Google even on the most current versions of Java, with version 2.11 of GWT released in 2024.

Our most complex transformation by far was the migration from on-premise operation to operating on Google Cloud Platform (GCP). In an 18 month project launched in 2019, we simultaneously moved from Oracle to a combination of PostgreSQL and Google BigQuery, and from operation on dedicated Unix VMs managed by our in-house sysadmin team to operation on a Kubernetes cluster in a Virtual Private Cloud (VPC) on Google Cloud Platform (GCP). The decision to split our data between two different sources was not taken lightly, as there was only one database in the original Oracle implementation. We found that clinical data was best managed in BigQuery due to its large size and relatively stable data footprint; we update the clinical dataset with EMR data once a day. All transactional application data such as user logins, saving queries and so on, is managed in a PostgreSQL Cloud SQL database because BigQuery is an analytic database and is not designed to work as a transactional database to support an application. An "external query" linkage between the two provides the glue between the two data sources. All clinical data queries are initiated on BigQuery, but many contain an external reference to the PostgreSQL database, as that is where per-study patient lists are stored.

Moving to Google BigQuery for our clinical data had a profound impact on our ability to revise our data mappings and respond to new customer needs.  Previously, some changes to our data model were practically infeasible, due to the time cost of reprocessing all of the historical data. BigQuery made it possible for us to reprocess the entire clinical dataset in hours instead of the days or even weeks required with our on-premise Oracle database. Table 1 attempts to compare the previous on-premise solution to the current cloud solution.

|  |  | On-Premise Solution | Cloud Solution (GCP) |
|---|---|---|---|
| Infrastructure | Hardware Config | Physical + Virtual Dell R830 (40 Cores, 1 TB RAM, 39 TB SSD) VMWare: 2 Virtual Servers (10 Cores, 36 GB RAM, 2 TB SAS) | n/a |
|  | SaaS Config | n/a | CloudSQL BigQuery (400-slot res) GKE |
|  | System Use | CPU: 40 Cores RAM: 1 TB Disk: 1-2 TB SSD | CloudSQL: 128 GB BigQuery: 2+ TB data BigQuery: 40 slot avg GKE: 6 Cores, 20 GB |
| Database | Application DB | Oracle DB | Cloud SQL (PostgreSQL) |
|  | Clinical DB |  | BigQuery |
|  | Compute Model | Physical Hardware | SaaS |
| Application | Language(s) | Java/GWT | Java/GWT |
|  | Compute Model | Virtual Servers (ESX) | Containerized (GKE) |
|  | Deployment Model | Weekly refresh of container by automated script | Automatic deployment of new container/config by Flux |
| Support | Application Support | In-house Dev team | In-house Dev team |
|  | DB Management | In-house DB team | n/a (SaaS) |
|  | Server Management | In-house Infrastructure team | n/a (SaaS) |
| Finance | Cost Model | CapEx (Hardware) + OpEx (Licenses + FTEs + Support) | OpEx (SaaS + FTEs) |

Table 1: On-Premise vs. Cloud Solution Comparison

## Features of Cohort Tool

Stanford researchers can search STARR to identify potential research patient cohorts using a self-service visual query tool called the Cohort Discovery Tool [Lowe2009] (see Figure 3). This tools relies on the data in Square Tables data model.

Figure 3: Cohort Discovery Tool running in a web browser

The original Cohort Discovery Tool only permitted searching for patients by demographics, date range, diagnosis, procedure, clinical document keyword search, lab results and medication orders. Based on valuable feedback from our user community, we have enhanced Cohort Discovery in the following ways:

- Patient age and time frame constraints can be applied individually to each clinical constraint, rather than only being available globally.

- Clinical constraints can be further modified by their occurrence frequency, e.g. only count patients with 2 or more occurrences of the specified condition.

- New search constraint types added: encounters with providers and/or in specified clinics, admissions by department, text found in nursing flowsheets, drug therapeutic class, patient language, vital status, the "Gene Pool" biobank, and previously saved cohorts.

- The clinical document constraint can now be further refined by specifying the clinical document type(s) in which to search.

- A powerful new temporal constraint allows combining two clinical constraints and specifying their temporal relationship, e.g. diagnosis of atrial fibrillation preceded administration of Coumadin by 1 or more days.

- Administrative features allow super-users to open another user's saved search and see the Structured Query Language (SQL) for the current search.

- Define a cohort by uploading a list of patient MRNs or radiology image accession numbers.

- Built-in support for research recruitment in the form of a downloadable report of patient names and detailed contact information for the current saved patient cohort. This feature is available only to Stanford's other "honest broker" intermediaries, not to the researchers.

## Features of Chart Review Tool

STARR Tools also provides a research Chart Review Tool (see Figure 4) designed to allow examination of clinical data to help determine if patients are suitable for inclusion in research studies. Chart Review Tool relies on the data in Square Tables data model. Each user can maintain a personal library of research patient cohorts within the system with access to PHI controlled by Institutional Review Board (IRB) approval.

The Chart Review Tool gives researchers online access to all clinical data in the STARR data model, including patient demographics, encounters, clinical documentation, lab results, diagnosis codes, procedures performed, medication order and administration records, and last known vital status. Additionally, if the researcher defined the cohort using a Cohort Discovery search, they can request that the chart review be kept up to date by periodic re-runs of the search; newly qualified patients are automatically added to chart review and they are notified by email.

Arguably the most important new feature we added was the capability to conduct self-service chart review and data download. The key enabler here is the sophisticated integration described in greater detail in the "Regulatory Compliance" section below between Stanford's eProtocol system operated by the Research Compliance Office (RCO) and the REDCap survey that houses the "Data Privacy Attestation" (DPA) survey designed and reviewed by Stanford's University Privacy Office (UPO). The process of obtaining approval from both the RCO and UPO places the research team on formal notice of their obligations in regards to data privacy, and raises awareness of the crucial importance of treating the data obtained via the self-service tools with utmost care for its confidentiality.

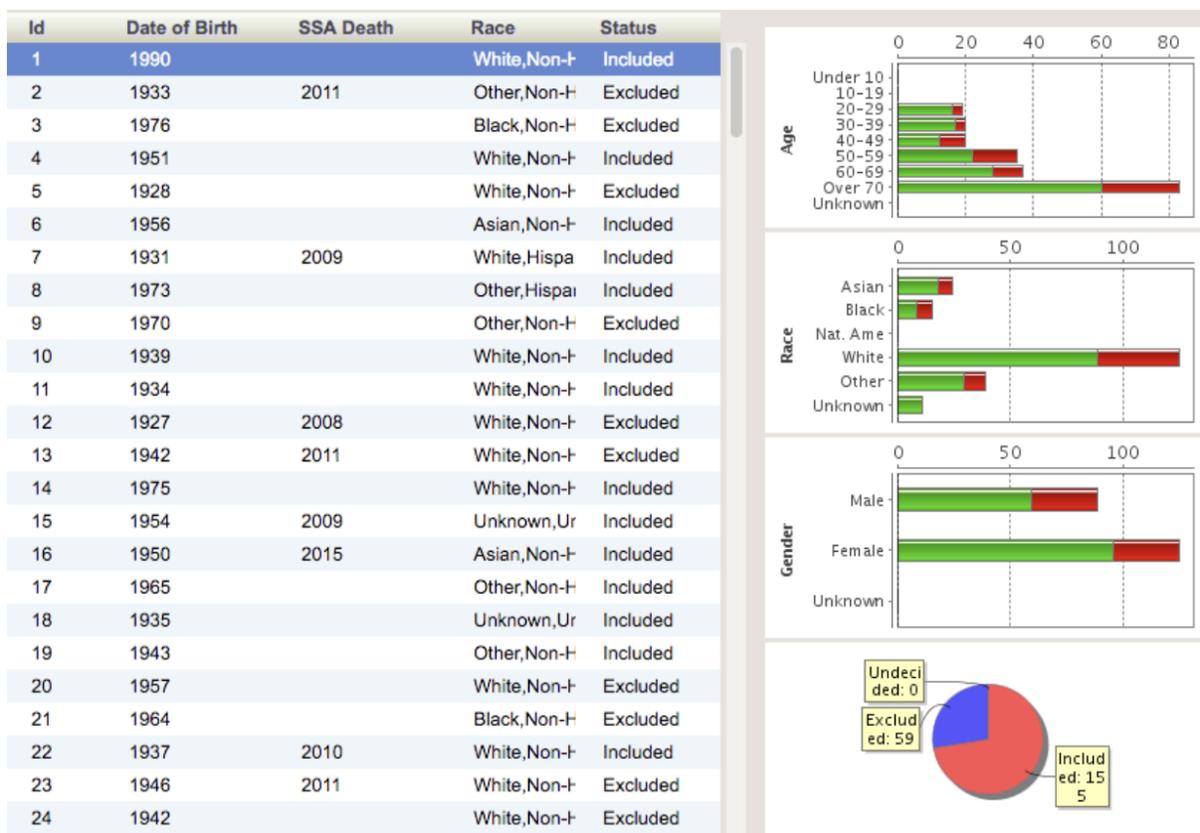

Figure 4: Chart Review Tool

## Data download tool

The data download tool, the latest feature, provides two separate actions: a) It allows the researcher to download data in the square tables data model or the OMOP data model, b) it allows for the data payload to be delivered to a researcher-owned end-point e.g. a secure laptop or Stanford's cloud-based computational platform for high-risk research data, Nero [Callahan2023, Datta2020].

## STARR suite of APIs

A key self-service enabler was the establishment of a secure API for identity verification, anonymization support, and compliance credential verification (see Figure 5). The identity verification API supports both REDCap and the Stanford Biobank. The anonymization support API is used not only by various software systems, specifically the Chart Review Tool as well as both DICOM image and OMOP CDM de-identification algorithms, but also extensively by the research consultation services team when creating custom datasets. The compliance verification API is used by REDCap and the Chart Review Tool.

Last but not least, we developed three APIs to support integration with other software systems and applications: identity verification, anonymization support, and compliance verification.

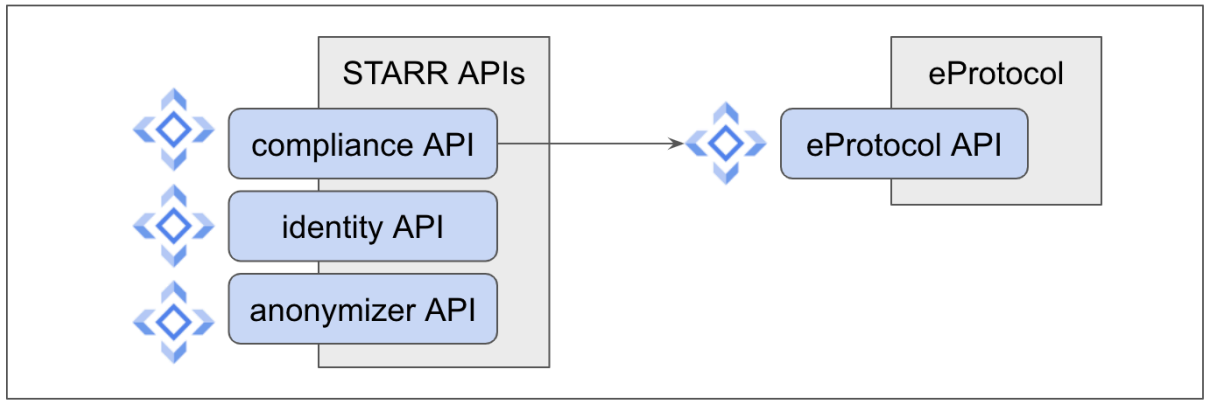

Figure 5: STARR APIs

The **identity verification API** expects a single MRN, and returns the associated patient name and date of birth if found. This is used by Stanford REDCap to implement a patient lookup mechanism for record creation, in order to ensure accurate data entry of this crucially important identifier. It is also used for a similar reason by the Stanford Biobank in their third-party biobanking management system.

The **anonymizer API** was developed to support scrubbing PHI from patient data for research as required by the "Minimum Necessary" principle in HIPAA. Scrubbing alphanumeric identifiers such as patient names and MRNs is achieved by creating a new identifier, unique to both the original identifier and the study, and storing it for re-use. The first time the API is called with a name or MRN, the API creates a new code; subsequent calls to the API with the same information can either return the same code returned previously, if identifier stability is required, or another randomly selected replacement, if the patient identity is already established and the goal is to obscure all identifiers. The system calling the API replaces the original identifier with the supplied coded identifier, thereby greatly reducing the risk of working with the data.

Scrubbing PHI is particularly complicated when it comes to dates. All dates pertaining to a patient's chart are considered identifying, but most research projects use and develop tools that expect to be working with patient timelines and actual dates. In accordance with recent standard practice [Hripcsak2016], the anonymization support API takes in a study identifier and a list of MRNs and returns a list of records consisting of the MRN paired with a stable pseudo-identifier that can be safely substituted for the MRN. Furthermore, after internal UPO review and risk assessment, Stanford decided to adopt the following approach to date masking. Each stable pseudo-identifier is accompanied by a stable but randomly generated non-zero number specific to the given study, which is standard practice, but Stanford uses a window between -30 and 30. We do this to support behavioral studies that rely on knowing the time of year.

The random number is used to shift dates by the system that invokes the API. The resulting "codebook" associated with the study is saved, and while it can be added to over time, a patient's record in the codebook does not change, which means that you always get back the same date "jitter" and the same patient coded id every time you call the API for a patient in a given study. It also means that all dates for a given patient will be shifted by the same amount, preserving the temporal relationships within that patient's history.

The **compliance API** returns a list of all known compliance documents associated with the supplied search term, which can be either a protocol number or a Stanford Single Sign On (SSO) identifier. The compliance API calls another API operated by the eProtocol team to look up IRB protocol data in real time. In an example, STARR Tools uses the compliance APIs to verify IRB status, researcher's status and other information made available via the IRB APIs.

API endpoint security is handled using a custom designed exchange of time-sensitive tokens in conjunction with an allow-list of trusted hosts.

## DevSecOps

As STARR Tools handles PHI, our application has always been classified by Stanford's Minimum Security Standards as High Risk PHI. The baseline for any application deployed at Stanford, whether or not it processes High Risk data, is to apply operating system security patches in a timely manner, conduct vulnerability scanning at least monthly, maintain up to date inventory records, use firewalls, and enforce password complexity. Low risk applications also need to require a 2nd factor for administrators, forward logs to a centrally managed log server, deploy malware and intrusion detection, and physically protect application servers. Additionally, high risk applications must access administrative accounts only through a secure endpoint, have conducted a formal Data Risk Assessment prior to deploying, and in our case demonstrate compliance with HIPAA export controls.

Several of these requirements, namely timely application of security patches to operating systems, regular vulnerability scanning, inventory management, intrusion detection, and physical security for server machines, are met by GCP itself, or by Stanford's integration with GCP. End users authenticate to the application through an Identity Aware Proxy that delegates authentication to the same SAML-based identity system used by on-premise applications. Two-factor authentication is required by that integration. Logs are maintained within the secure GCP environment, as they may contain PHI that is not allowed in the centralized logging service used for on-premise applications.

All of the clinical data stored in Google Cloud is protected by custom rules using Google's VPC Service Controls feature. These rules define our Service Perimeter, which, like a firewall, defines an extra layer of protection about the identities and endpoints that are permitted to access resources in our cloud environment. Google VPC also offers infrastructure-level access controls, which we use to limit access to the back-end services such as BigQuery to a small group of developers and data scientists. We also impose application-specific controls on access to the data by end users, discussed in greater detail in the "Regulatory Compliance" section below.

Stanford uses Wiz (https://www.wiz.io/), a security scanning tool, to look for vulnerabilities including misconfigurations across all of its cloud projects. We are presently engaged in an effort to reduce alert fatigue by adjusting the templates to take into account known deployment patterns. Wiz scans are not limited to virtual machines; it examines many cloud service configuration issues, too. In addition, the cloud ops team conducts a quarterly review of all changes and current configurations, paying particular attention to BigQuery and other persistence layer services that are outside the scope of Wiz monitoring.

In addition to meeting or exceeding the minimum security standards for software deployment, we also go to great lengths to ensure application security at the code layer. We use Sonar Cloud for static code analysis and code coverage and Snyk for code vulnerability scanning of imported libraries.

The new application deployment setup involves continuous integration and deployment (CI/CD) with source code maintained in GitHub, Travis-CI for building and deploying new images, and Flux triggers for automated continuous deployment of the container images to a standard Google Kubernetes Engine (GKE) cluster within Stanford's GCP VPC.

## Regulatory Compliance

STARR is operated as a Stanford IRB-approved research protocol, with a waiver of informed consent and HIPAA authorization, that permits all clinical data captured as part of routine patient care at Stanford Medicine to be transferred to STARR. The transfer of clinical data to STARR is further governed by a detailed agreement between Stanford's two hospitals and the Stanford University School of Medicine, where the project is based. STARR IRB is reviewed and renewed annually and the PI is the Senior Associate Dean of Research at School of Medicine or their faculty delegate.

When first launched, the only self-service tool was Cohort Discovery, which only exposes counts and no actual patient data.  To achieve self-service chart review and data download while maintaining acceptably low levels of risk of inadvertent disclosure, we embarked on a multi-year partnership with both the Stanford University Privacy Office (UPO, https://privacy.stanford.edu/), the team responsible for protecting data confidentiality, and the Stanford Research Compliance Office (RCO, https://researchcompliance.stanford.edu/), the team responsible for overseeing the ethics of human subjects research.

In 2017, we developed an integration with eProtocol, the system used by the RCO to track all Stanford research protocols.  The user clicks on a link in the header of the "Confidentiality Protections" section of eProtocol and is taken to the Data Privacy Attestation (DPA) REDCap survey (Figure 6), which when launched in this manner is pre-filled with information about the currently open modification of the protocol. The survey asks detailed questions about what PHI will be used and for what purpose (recruitment, internal use, or external disclosure), the categories of clinical data, and whether any data will cross international borders. Finally, the researcher attests to a set of statements about data use and handling. They then type in their name by way of signing the document, and save it.

Once saved, the system sends a summary of the completed survey back to the originating browser where it is picked up by custom Javascript in eProtocol and inserted in read-only form into the protocol document.  If an attempt is made to launch the DPA survey from any other context, the user will see an error message. Research project DPAs must be created from within the context of eProtocol, to enable the joint review of the information by Privacy and the RCO. The Privacy Office is notified when a new DPA is submitted, and cross-checks that the PHI and clinical data types make sense for the research purposes stated in the associated protocol. They can and do reject applications that seem out of line with the stated research goals. Once the researcher comes to an agreement with Privacy in regard to the scope of PHI and clinical data permitted for the project, the protocol is ready for approval by the IRB panel.

Once this system was completed and verified to work reliably, we finally felt ready to take the step of allowing self-service access to patient charts.  We wrote up comprehensive documentation on how to complete the compliance approvals process, and announced self-service chart review availability in November 2017. To review charts online, the researcher must have supplied a currently valid DPA when saving the cohort for review. That being said, the compliance ecosystem that has grown up around STARR Tools is surprisingly complex, as evinced by the depth of the online documentation.

Figure 6: Screenshot of the output of our protocol/DPA validation tool

Data that has been scrubbed of PHI is readily available, however, at the time of writing, the PHI scrubbed data is not deemed de-identified by the UPO. The data is deemed Moderate or High Risk and depends on the specifics of the data type. For instance, presence of PHI scrubbed unstructured data such as clinical text or DICOMs are deemed High Risk data. Stanford University computational and storage systems that meet minimum security requirements for the Moderate Risk and High Risk data can be used to store and analyze the downloaded data. If the researcher wishes to move the Moderate or High Risk data elsewhere, they need to file for a Data Risk Assessment (DRA, https://uit.stanford.edu/security/dra) with the University and the DRA team makes an expert determination regarding the risk mitigation strategies.

If PHI download is requested by the researcher, additional human-in-the-loop interviews may be required. All downloaded files have word "confidential" appended to the file name to further remind the user regarding the data risk. In the case of an eventuality, STARR Tools include robust logging framework and audit capabilities.

In rare cases, STARR Tools support non-human subject research such as education or clinical quality improvement. In such cases, a standalone DPA is available to user. These workflows may require additional human-in-the-loop review or hospital sponsorship.

## Results

The STARR Tools system has seen steady growth in popularity over the years, as shown in Figure 7 (number of distinct users) and Figure 8 (number of projects/protocols). Figure 9 shows the median cohort size. *The 2024 numbers are presented through end of Nov 2024.*

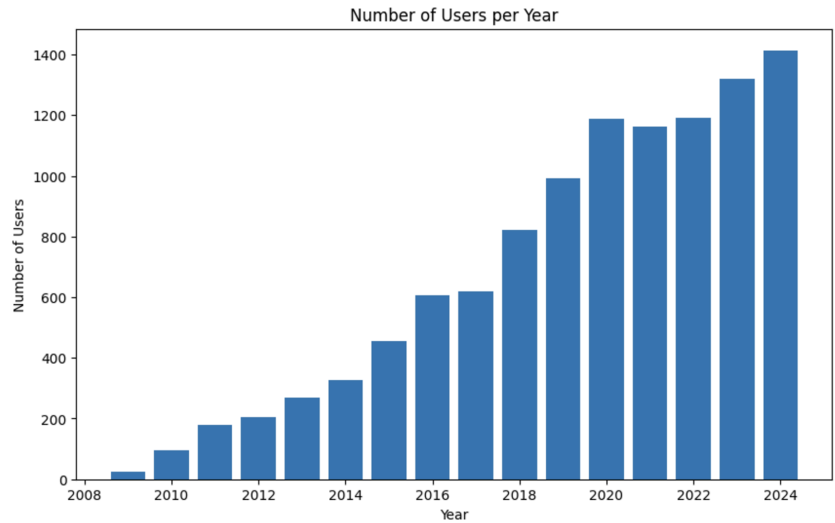

Figure 7: STARR Tools: number of distinct users per year

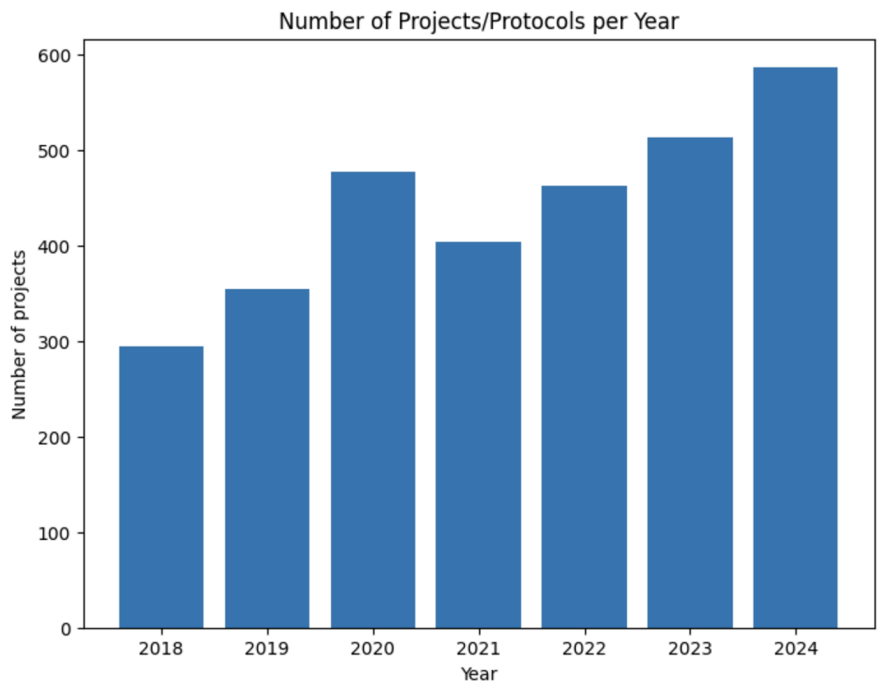

Figure 8: STARR Tools: number of distinct research projects supported per year

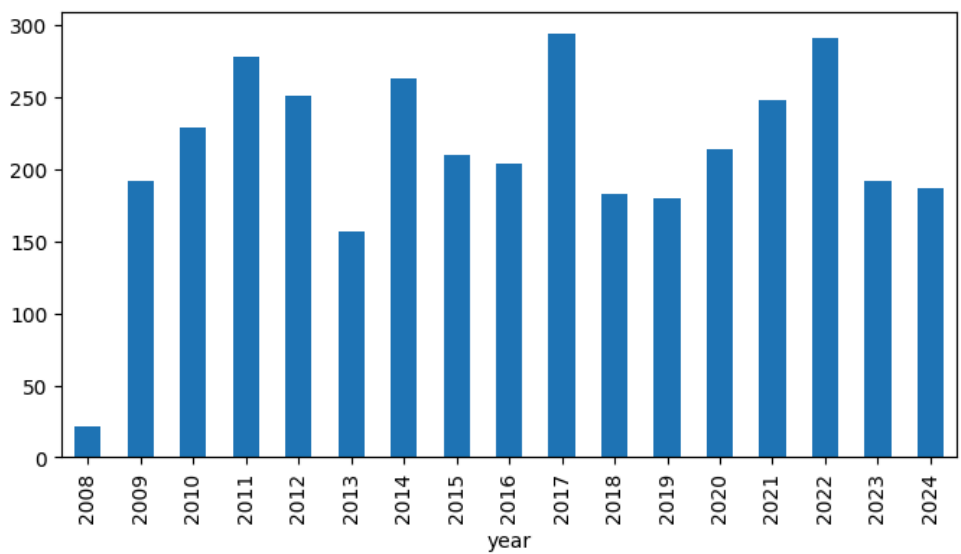

Figure 9: STARR Tools: Median size of saved cohorts

## Discussion

From its early roots as a standards-based enterprise Informatics model supporting Clinical and Translational Research, STARR Tools has matured into a system that provides Stanford researchers rich, flexible and HIPAA-compliant self-service access to a wide range of clinical data for research purposes. STARR Tools has evolved to keep up with the changes in hospital EHRs (e.g. Legacy to Epic, Cerner to Epic), changes in data (e.g. introduction of Epic Beacon and Beaker modules), changes in infrastructure (e.g. on-premise to cloud) and changes to regulatory requirements. The system today provides not only sophisticated cohort discovery and in depth chart review and data download via self-service web applications, but also supports an extensive set of APIs for data integration with other systems such as REDCap, Biobank and for PHI scrubbing. STARR Tools is poised to offer download to approved computational environment such as Nero, download of EMR data in OHDSI Observational Medical Outcomes Partnership (OMOP) Common Data Model (CDM) and download of other data modalities such as radiology DICOMs.

## Acknowledgement

Using CRediT taxonomy(Contributor Roles Taxonomy, https://credit.niso.org/), we present the contributing roles for our authors - SC Weber (Conceptualization, Methodology, Supervision, Project administration, Writing - original draft), J Pallas (Software, Formal Analysis, Validation, Visualization, Writing - original draft), G Olson (Methodology), D Love (Formal Analysis), S Malunjkar (Software), S Boosi (Conceptualization, Software, Supervision, Project administration), E Loh (Software), S Datta (Resources, Writing - review and editing), TA Ferris (Conceptualization, Funding acquisition)

STARR suite (2008-), including the first generation STRIDE, are made possible by Stanford School of Medicine Dean's Office.